\documentclass[final,3p,times]{elsarticle} 

\usepackage{amsmath,amssymb,amsthm}
\usepackage{bm,multirow,xcolor,colortbl}
\usepackage{subfig}
\usepackage{algorithm,algorithmic}
\usepackage[capitalize,noabbrev]{cleveref}

\theoremstyle{definition}
\newtheorem{property}{Property}
\crefname{property}{Property}{Properties}

\numberwithin{equation}{section}

\newcommand{\gap}{\hspace{0.1cm}}

\newcommand{\hD}{\widehat{D}}

\newcommand{\bA}{\mathbf{A}}
\newcommand{\bJ}{\mathbf{J}}

\newcommand{\bP}{\mathbf{P}}
\newcommand{\bb}{\mathbf{b}}
\newcommand{\brho}{\boldsymbol{\rho}}

\newcommand{\depsilon}{\Delta \epsilon}
\newcommand{\dalpha}{\Delta \alpha}
\newcommand{\dd}{\Delta d}
\newcommand{\dtheta}{\Delta \theta}

\DeclareMathOperator{\diag}{diag}

\definecolor{Gray}{gray}{0.9}

\usepackage{tikz}
\usetikzlibrary{arrows.meta,shapes.geometric}
\tikzset{%
  >={Latex[width=2mm,length=2mm]},
         block/.style = {rectangle, rounded corners, draw=black, text centered, minimum width=3.5cm, minimum height=1cm},
         grayblock/.style = {block, fill=black!10},
         nullblock/.style = {text centered, minimum height=0.5cm},
         ifblock/.style = {diamond, aspect=2.5, draw=black, text centered, minimum width=3.5cm, minimum height=1cm}
}

\journal{arXiv}

\begin{document}

\begin{frontmatter}
\title{A numerically efficient output-only system-identification framework for stochastically forced self-sustained oscillators}

\author[ML]{Minwoo Lee}
\address[ML]{Department of Mechanical Engineering, Hanbat National University, Daejeon 34158, Korea}
\ead{mwlee@hanbat.ac.kr}

\author[KTK]{Kyu Tae Kim}
\address[KTK]{Department of Aerospace Engineering, Korea Advanced Institute of Science and Technology~(KAIST), Daejeon 34141, Korea}
\ead{kt_kim@kaist.ac.kr}

\author[JP]{Jongho Park\corref{cor}}
\ead{jongho.park@kaust.edu.sa}
\ead[url]{https://sites.google.com/view/jonghopark}
\address[JP]{Computer, Electrical and Mathematical Science and Engineering Division, King Abdullah University of Science and Technology~(KAUST), Thuwal 23955, Saudi Arabia}

\cortext[cor]{Corresponding author}

\begin{abstract}
Self-sustained oscillations are ubiquitous in nature and engineering. In this paper, we propose a novel output-only system-identification framework for identifying the system parameters of a self-sustained oscillator affected by Gaussian white noise. A Langevin model that characterizes the self-sustained oscillator is postulated, and the corresponding Fokker--Planck equation is derived from stochastic averaging. From the drift and diffusion terms of the Fokker--Planck equation, unknown parameters of the system are identified. We develop a numerically efficient algorithm for enhancing the accuracy of parameter identification. In particular, a modified Levenberg--Marquardt optimization algorithm tailored to output-only system identification is introduced. The proposed framework is demonstrated on both numerical and experimental oscillators with varying system parameters that develop into self-sustained oscillations. The results show that the computational cost required for performing the system identification is dramatically reduced by using the proposed framework. Also, system parameters that were difficult to be extracted with the existing method could be efficiently computed with the system identification method developed in this study. Pertaining to the robustness and computational efficiency of the presented framework, this study can contribute to an accurate and fast diagnosis of dynamical systems under stochastic forcing.
\end{abstract}

%%Research highlights
%\begin{highlights}
%\item TO BE MODIFIED
%\end{highlights}

\end{frontmatter}

%% \linenumbers

% Section: Introduction
\section{Introduction}
\label{Sec:Introduction}

A self-sustained oscillation, also known as self-excited oscillation or autonomous oscillation, in a nonlinear system occurs from a balance between the driving energy source and the damping mechanism~\cite{pikovsky2003,son2023}. In a noise-free environment, the amplitude of a self-sustained oscillation does not grow or decay in time, while the dynamics of such an oscillation is governed by the oscillator itself, rather than the initial conditions~\cite{balanov2008}. Self-sustained oscillation is found ubiquitously in many natural and engineered systems, including heartbeat~\cite{jenkins2013}, stellar pulsations~\cite{fuller2017}, hydrodynamic~\cite{zhu2017onset,lee_2019}, thermoacoustic~\cite{guan2019open,lee2020} and aeroacoustic~\cite{boujo2020processing} oscillations, to name just a few. 

Over the past few decades, much effort has been devoted to understanding and describing the dynamics of self-sustained oscillations. This is because, in practical engineered systems, vibrations characterized by self-sustained oscillation are often detrimental. For instance, self-sustained thermoacoustic oscillations that are induced by positive feedback between pressure and heat release rate oscillation in a combustor can induce vibration and fatigue, decreasing the lifetime of the overall system~\cite{Lieuwen2005}. It is therefore important to be able to understand the characteristics of self-sustained oscillation, so as to predict, avoid or control such an oscillation. 

A promising method for understanding the dynamics of a self-sustained oscillator is the system identification approach, which is a type of inverse problem. In a system identification approach, a low-order model is postulated to describe the system's dynamics\footnote{System model can sometimes be identified from the system-identification process itself~\cite{NEGRINI}.}, and the parameters of that model are identified~\cite{boujo2020processing,Karch1995,Breccolotti2008}. Two classes of the system-identification framework are present---input-output and output-only. In the input-output framework~\cite{lee_2019,lee2020,LIU20102615,POLIFKE2014109,SOVARDI201690}, system's response to the input signal is used for system identification. In this framework, one can adjust the intensity of the forcing (i.e., the input signal), so as to induce the desired response of the system that is optimal for system identification. For example, coherence resonance~\cite{pikovsky1997coherence,ushakov2005coherence} that occurs at an intermediate intensity of noise input can be artificially induced to perform the system identification~\cite{lee_2019}. However, it is often difficult or impractical to apply external forcing in many systems. High-pressure oscillations occurring in rocket combustors and gas turbines, for example, are difficult to be perturbed by external forcing owing to the high energy required. In such cases, the output-only framework~\cite{Yu2008, NOIRAY2013152, BN:2017, LKGL:2021} that relies solely on the output signal of the system should be applied. In this framework, observable data inherent to the system is processed using various tools, such as modal identification~\cite{nagarajaiah2009}, autosynchronization~\cite{Yu2008} and sparse regression~\cite{RABK:2019}, for finding the unknown parameters of the system.

A notable approach in output-only system identification of self-sustained oscillators utilizes the effect of random fluctuations acting on the system. Unlike the conventional viewpoint that noise is considered as signal contamination, this approach focuses on the dynamic effect of noise that drives the oscillator away from its original dynamical state. For this purpose, statistical methods based on the Langevin equation and the corresponding Fokker--Planck equation are used to extract the deterministic information from the stochastic signal. This type of output-only system-identification framework has been demonstrated in various systems, both numerically and experimentally~\cite{boujo2020processing,NOIRAY2013152,BN:2017,LKGL:2021,bonciolini2017,lee2023data}.

One of the main challenges when applying the above-mentioned system-identification framework is to overcome the adverse \textit{finite-time effect}. This effect is caused by the non-Markovian nature of the noise, owing to the coarse sampling rate or the band-pass filtering around the major oscillation frequency, notwithstanding the Markovian assumption of memoryless noise. An efficient method for addressing the finite-time effect is to use the adjoint Fokker--Planck equation for estimating the deterministic parameters of the system~\cite{lade2009,HF:2011}. Specifically, Boujo and Noiray~\cite{BN:2017} showed that coefficients of Van der Pol-type self-sustained oscillator can be accurately extracted via optimization based on adjoint Fokker--Planck equation. In their study, a direct search method, namely the Nelder--Mead optimization~\cite{LRWW:1998,LPW:2012}, is performed to minimize the difference between experimental and mathematical drift/diffusion terms, identifying the parameters that best represent the dynamics of self-sustained oscillation. In a more recent study, this framework is experimentally validated on an aeroacoustic system~\cite{BN:2019}.

In this study, we build on the previous work by~\cite{BN:2017}, focusing on the numerical efficiency of the system identification. Although the established adjoint-based system identification method provides sufficiently accurate system parameters, it requires much computational cost for optimization. This is because the convergence rate of direct search methods can be very slow~\cite{LTT:2000}; direct search methods do not utilize the first- or higher-order derivatives of the cost function to improve the convergence rate. Considering the practical applications, however, it is important that the system identification is conducted quickly. This is because, in many systems~(e.g., gas turbines or rocket combustors), it is often essential to perform instantaneous preventive or circumventive measures for self-sustained oscillation, before such oscillation causes severe system damage.

Addressing this issue, we further develop the adjoint-based output-only system-identification framework by improving its numerical aspect. We first observe that the optimization model used in output-only system identification is a nonlinear least-squares problem. Although Gauss--Newton-type algorithms such as the Levenberg--Marquardt algorithm~\cite{Marquardt:1963} are known to be good choices to ensure fast convergence for nonlinear least-squares problems~\cite{Yuan:2011}, we cannot directly apply them to our optimization model because it is difficult to compute the first-order derivatives of the cost function. For the sake of achieving the convergence rate comparable to Gauss--Newton-type methods without the information of the first-order derivatives, we propose a novel optimization algorithm for the output-only system identification. We construct the optimization algorithm by replacing the Jacobian terms in the Levenberg--Marquardt algorithm by their suitable finite difference approximations. Since the proposed algorithm can efficiently utilize an approximated derivative information of the cost function to find a good search direction, we hypothesize that such an algorithm can find an optimum of the system identification model along a more optimized trajectory, compared to the conventional Nelder--Mead algorithm. Thus, it is expected that the proposed algorithm requires a much smaller number of residual computations than the Nelder--Mead algorithm, implying that the computational cost of the proposed algorithm can be dramatically improved. From the proposed approach, we aim to lessen the computational cost required for the optimization, enhancing the overall efficiency of the system-identification framework.

This paper is organized as follows. In \cref{Sec:SI}, we review the established system-identification framework of stochastically forced self-sustained oscillators and relevant existing numerical algorithms. In \cref{Sec:Proposed}, we propose a new numerical algorithm for solving the optimization model for system identification. Numerical results and gas-turbine combustor experiments that highlight the robustness of the proposed algorithm are presented in \cref{Sec:Demonstration}. We conclude the paper with remarks in \cref{Sec:Conclusion}.

\section{System-identification framework}
\label{Sec:SI}
In this section, we review the system-identification framework for stochastically forced self-sustained oscillators presented in~\cite{BN:2017, LKGL:2021}. We first present the Fokker--Planck equation that is derived from the Langevin model characterizing the self-sustained oscillator under random forcing. Then we review the optimization-based output-only system-identification method~\cite{BN:2017,LKGL:2021} that enables the extraction of system parameters from the drift and diffusion terms of the Fokker--Planck equation.

% Subsection: FP
\subsection{Fokker--Planck equation}
We consider a phenomenological model for a self-sustained oscillator under stochastic forcing~\cite{son2023, lee2023numerical}. Specifically, a Langevin equation characterizing a Van der Pol-type oscillator perturbed by additive white Gaussian noise is postulated:
\begin{equation}
\label{VdP}
\frac{d^2 x}{dt^2} - \left( \epsilon + \alpha x^2 \right) \frac{dx}{dt} + \omega^2 x = \sqrt{2d} \eta,
\quad t > 0,
\end{equation}
where $x(t)$ is the instantaneous state of the system, $\eta (t)$ is a unit white Gaussian noise term, $d$ is the noise amplitude, $\omega$ is the angular frequency, $\epsilon$ is the linear growth~(positive) or decay~(negative) coefficient, and $\alpha$ is the negative nonlinear coefficient. Specifically, the system is in the linearly stable regime at $\epsilon<0$, crosses the Hopf point at $\epsilon=0$, and develops a self-sustained oscillation in the linearly unstable regime at $\epsilon>0$. In this paper, we not only consider the self-sustained oscillation at $\epsilon>0$ but also consider the negative or zero $\epsilon$ cases, as the system identification at the latter cases can lead to a prediction of self-sustained oscillations \cite{lee_phd}.

To find a probabilistic solution of~\eqref{VdP}, we consider the following Fokker--Planck equation, which is obtained by stochastic averaging of~\eqref{VdP}~(see \ref{App:FP}):
\begin{equation} \begin{split}
\label{FP}
&\frac{\partial}{\partial t} P (a,t) = - \frac{\partial}{\partial a} \left[ D^{(1)}(a) P(a,t) \right] + \frac{\partial^2}{\partial a^2} \left[ D^{(2)}(a) P(a,t) \right],
\quad a > 0, \gap t > 0, \\
&P(a,0) = P_0 (a), \quad a > 0, \\
&P(0, t) = 0, \quad t > 0,
\end{split} \end{equation}
where $P(a,t)$ is the transitional probability density function of $a$ at time $t$, $D^{(1)}(a)$ and $D^{(2)}(a)$ are the Kramer--Moyal coefficients, and $P_0 (a)$ is the initial probability density at $t=0$. Specifically, $D^{(1)}(a)$ and $D^{(2)}(a)$ represent the drift and diffusion terms, respectively, and are given by
\begin{equation}
\label{KMc}
D^{(1)}(a) = \frac{\epsilon}{2} a + \frac{\alpha}{8} a^3 + \frac{d}{2\omega^2 a}, \quad
D^{(2)}(a) = \frac{d}{2\omega^2}.
\end{equation}
%We note that, although $D^{(2)}(a)$ is in fact independent of $a$, we use this notation in the remainder of this paper for consistency and convenience.

% Subsection: SI model
\subsection{System identification model}
Among the unknown parameters of the Langevin equation~\eqref{VdP}, we can easily identify $\omega$ from spectral analysis, specifically by inspecting the power spectral density.
Although the noise may induce a shift in $\omega$ due to the anisochronicity factor~\cite{Zakh2010}, for simplicity, we assume that $\omega$ is independent of $d$, as in~\cite{lee2020}.
%Although the noise can cause the shift of $\omega$ due to the anisochronicity factor~\cite{Zakh2010}, we assume that $\omega$ is independent of $d$, for the sake of mathematical simplicity. This assumption is found to be valid in some previous studies regarding an experimental thermoacoustic system~\cite{lee2020}.
Finding other system parameters, namely $\epsilon$, $\alpha$ and $d$, is a more challenging task.
While the most straightforward way to identify these parameters using the observed instantaneous state $x(t)$ is the extrapolation method, which is described in \ref{App:Extrapolation}, it is known to be prone to the adverse effect of the finite-time effect.
Here, we summarize key features of the optimization model for system identification proposed in~\cite{BN:2017,LKGL:2021}, which alleviates the adverse finite-time effect.

The starting point is the Fokker--Planck equation~\eqref{FP}.
In order to extract the system parameters $\epsilon$, $\alpha$, and $d$ of~\eqref{VdP}, the drift~($D^{(1)}(a)$) and diffusion~($D^{(2)}(a)$) terms in~\eqref{FP} have to be first identified. Siegert et al.~\cite{Siegert1998} have shown that, these terms can be computed from the time correlation of the output signal. 
More precisely, for $n=1,2$ and $a > 0$, we have
\begin{equation}
\label{D_tau_limit}
D^{(n)}(a) = \lim_{\tau \rightarrow 0} D_{\tau}^{(n)}(a),
\end{equation}
where $D_{\tau}^{(n)}(a)$ is defined by
\begin{equation}
\label{D_tau_exact}
D_{\tau}^{(n)}(a) = \frac{1}{n! \tau} \int_0^{\infty} (A-a)^n P(A, t + \tau \mid a, t) \,dA.
\end{equation}
In~\eqref{D_tau_exact}, $P(A, t+\tau \mid a,t)$ denotes the conditional probability density function of the oscillation amplitude being $A$ at time $t + \tau$ in case where the amplitude is $a$ at time $t$. Using the identity~\eqref{D_tau_exact}, one can obtain the approximate value of $D_{\tau}^{(n)}(a)$ without analytically solving the Fokker--Planck equation~\eqref{FP}.

Assume that the system state $x(t)$ is available on the time interval $t \in [0, T_{\max}]$ for some $T_{\max} > 0$.
We choose a set of time-shifts $\{ \tau_j \}_{j=1}^{N_{\tau}} $ on the time interval for some positive integer $N_{\tau}$.
Applying the Hilbert transform, amplitude $a(t)$ can be extracted from $x(t)$.
We can then choose a set of amplitudes $\{ a_i \}_{i=1}^{N_a} $ in the range of $a(t)$ for some positive integer $N_a$.
%such that $a_{\min} \leq a_i \leq a_{\max}$ for all $i$, where
%\begin{equation}
%\label{a_range}
%a_{\min} = \min_{0 \leq j \leq N_t} a(t_j), \quad
%a_{\max} = \max_{0 \leq j \leq N_t} a(t_j).
%\end{equation}
In the remainder of this section, let the indices $i$ and $j$ run from $1$ to $N_a$ and $N_{\tau}$, respectively.
Each point $(a_i, \tau_j)$ will play a role of a sampling point in the amplitude-time space for system identification.
A discrete approximation of the conditional probability density function $P(A,t+\tau \mid a, t)$ in~\eqref{D_tau_exact} can be computed on the grid consisting of the points $(a_i, \tau_j)$ using the amplitude data $a(t)$.
Then the integral in~\eqref{D_tau_exact} can be evaluated by a numerical integration scheme such as the simple trapezoidal rule. In what follows, we denote the value of $D_{\tau}^{(n)}(a)$ at $a = a_i$ and $\tau = \tau_j$ computed from the system state $x(t)$ by the above-mentioned procedure as $\hD_{\tau_j}^{(n)}(a_i)$. That is, $\hD_{\tau_j}^{(n)}(a_i)$ depends on $x(t)$ only, so that a priori knowledge on the parameters $\epsilon$, $\alpha$, and $d$ in~\eqref{VdP} is not required in the computation of $\hD_{\tau_j}^{(n)}(a_i)$.

In the optimization model proposed in~\cite{BN:2017,LKGL:2021}, we minimize the $\ell^2$-difference between the experimental data $\hD_{\tau_j}^{(n)}(a_i)$ and the coefficient-based data $D_{\tau_j}^{(n)}(a_i)$. Here, the latter value should be calculated from a solution of the Fokker--Planck equation~\eqref{FP}. However, it is generally difficult to obtain a solution to the standard Fokker--Planck equation~\eqref{FP}, either analytically or numerically~\cite{DiPaola2002}. Therefore, we alternatively consider the following \textit{adjoint} Fokker--Planck equation with the system coefficients $\epsilon$, $\alpha$ and $d$~\cite{BN:2017,lade2009}:
%and compute an approximate solution to~\eqref{FP} \cite{BN:2017,lade2009}. Specifically, the adjoint Fokker--Planck equation is written as:
\begin{equation}
\label{AFP}
\begin{split}
&\frac{\partial}{\partial t} P^{\dag} (A, t) = D^{(1)}(A) \frac{\partial}{\partial A} P^{\dag} (A, t) + D^{(2)}(A) \frac{\partial^2}{\partial A^2} P^{\dag}(A, t),
\quad A > 0, \gap t > 0,\\
&P^{\dag}(A,0) = (A-a)^n, \quad A > 0, \\
&P^{\dag}(0, t) = 0, \quad t > 0,
\end{split}
\end{equation}
where $D^{(1)}(A)$ and $D^{(2)}(A)$ are given in~\eqref{KMc}.
%In general, an analytic solution of the time-dependent problem~\eqref{AFP} is not obtainable; see~\cite{HF:2011} for some special cases when the general solution can be obtained. 
We solve~\eqref{AFP} numerically, yielding an approximate $D_{\tau_j}^{(n)}(a_i)$ value~\cite{lade2009}:
\begin{equation}
\label{D_tau_AFP}
D_{\tau_j}^{(n)} (a_i) = \frac{P^{\dag}(a_i, \tau_j)}{n! \tau}.
\end{equation}

Now, we consider the following cost function that measures an weighted $\ell^2$-error between $\hD_{\tau_j}^{(n)}(a_i)$ and $D_{\tau_j}^{(n)}(a_i)$ when the parameters $\epsilon$, $\alpha$, and $d$ in~\eqref{VdP} vary:
\begin{equation}
\label{cost}
E(\epsilon, \alpha, d) = \frac{1}{2N_a N_{\tau}} \| \brho ( \epsilon, \alpha, d ) \|_{\bP}^2
= \frac{1}{2N_{a} N_{\tau}} \sum_{n=1}^2 \sum_{i=1}^{N_a} \sum_{j=1}^{N_{\tau}}
P_{ij} \left( \hD_{\tau_j}^{(n)}(a_i) - D_{\tau_j}^{(n)}(a_i; \epsilon, \alpha, d) \right)^2,
\end{equation}
where $\bP \in \mathbb{R}^{2N_a N_{\tau} \times 2N_a N_{\tau}}$ is a diagonal matrix, with each entry representing the experimental probability density $P_{ij}$ measured at $a = a_i$ and $\tau = \tau_j$,
and $\rho (\epsilon, \alpha, d) \in \mathbb{R}^{2N_a N_{\tau}}$ is a residual vector whose entries are $\hD_{\tau_j}^{(n)}(a_i) - D_{\tau_j}^{(n)}(a_i; \epsilon, \alpha, d)$.
% \begin{equation}
% \label{cost}
% E(\epsilon, \alpha, d) = \frac{1}{2N_{a} N_{\tau}} \sum_{n=1}^2 \sum_{i=1}^{N_a} \sum_{j=1}^{N_{\tau}}
% P_{ij} \left( \hD_{\tau_j}^{(n)}(a_i) - D_{\tau_j}^{(n)}(a_i; \epsilon, \alpha, d) \right)^2,
% \end{equation}
% where $P_{ij}$ is the experimental probability density function measured at $a= a_i$ and $\tau= \tau_j$. In~\eqref{cost}, the notation $D_{\tau_j}^{(n)}(a_i; \epsilon, \alpha, d)$ is used for the sake of highlighting the dependency of $D_{\tau_j}^{(n)}(a_i)$ on the parameters $\epsilon$, $\alpha$, and $d$. Equivalently, we may write $E(\epsilon, \alpha, d)$ in a more compact manner as follows:
% \begin{equation*}
% E(\epsilon, \alpha, d) = \frac{1}{2N_a N_{\tau}} \| \brho ( \epsilon, \alpha, d ) \|_{\bP}^2,
% \end{equation*}
% where $\brho (\epsilon, \alpha, d)$ is a residual vector in $\mathbb{R}^{2N_a N_{\tau}}$ whose entry corresponding to the index $(n, i, j)$~($1 \leq n \leq 2$, $1 \leq i \leq N_a$, $1 \leq j \leq N_{\tau}$) is $\hD_{\tau_j}^{(n)}(a_i) - D_{\tau_j}^{(n)}(a_i; \epsilon, \alpha, d)$,
% $\bP$ is a diagonal matrix of size $2N_a N_{\tau} \times 2N_a N_{\tau}$ whose entry corresponding to the index $(n, i, j)$ is $P_{ij}$,
% and $\| \cdot \|_{\bP}$ denotes the weighted norm in $\mathbb{R}^{2 N_a N_{\tau}}$ induced by the matrix $\bP$.
Clearly, the core step in the computation of $E(\epsilon, \alpha, d)$ is an evaluation of the residual $\brho (\epsilon, \alpha, d)$.
The evaluation procedure of $\brho (\epsilon, \alpha, d)$ using the equations~\eqref{AFP} and~\eqref{D_tau_AFP} is summarized in \cref{Alg:residual}.

% Algorithm: Evaluation of the residual
\begin{algorithm}
\caption{Evaluation of the residual $\brho(\epsilon, \alpha, d)$}
\begin{algorithmic}[]
\label{Alg:residual}
\STATE \textsc{Inputs}: parameters $\epsilon$, $\alpha$, and $d$, estimates of the finite-time Kramer--Moyal coefficients $\{ \hD_{\tau_j}^{(n)}(a_i) : 1 \leq n \leq 2, \gap 1 \leq i \leq N_a, \gap 1 \leq j \leq N_{\tau} \}$%, probability values $\{ P_{ij} : 1 \leq i \leq N_a, \gap 1 \leq j \leq N_{\tau} \}$
\STATE \textsc{Outputs}: residual vector $\brho(\epsilon, \alpha, d)$

\STATE $\bullet$ Solve the adjoint Fokker--Planck equation~\eqref{AFP}. \quad ($1 \leq n \leq 2$, $1 \leq i \leq N_a$, $1 \leq j \leq N_{\tau}$)
\STATE $\bullet$ Compute $D_{\tau_j}^{(n)}(a_i)$ by invoking the formula~\eqref{D_tau_AFP}. \quad ($1 \leq n \leq 2$, $1 \leq i \leq N_a$, $1 \leq j \leq N_{\tau}$)
\STATE $\bullet$ Compute each entry $\hD_{\tau_j}^{(n)}(a_i) - D_{\tau_j}^{(n)}(a_i; \epsilon, \alpha, d)$ of $\brho(\epsilon, \alpha, d)$. \quad ($1 \leq n \leq 2$, $1 \leq i \leq N_a$, $1 \leq j \leq N_{\tau}$)
\end{algorithmic}
\end{algorithm}

Finally, the optimization of the system coefficients is done by solving the following minimization problem:
\begin{equation}
\label{min}
\min_{\epsilon,\alpha, d} E(\epsilon, \alpha, d).
\end{equation}
It was demonstrated mathematically in~\cite{LP:2023} that $\ell^2$-minimization models such as~\eqref{min} are robust to additive measurement noise.
In~\cite{BN:2017,LKGL:2021}, the Nelder--Mead algorithm~\cite{LRWW:1998} was adopted to solve~\eqref{min}. In the next section, on the contrary, we take a further step and propose a novel numerical algorithm for more efficiently solving~\eqref{min}. 

% Section: Proposed algorithm
\section{Proposed algorithm}
\label{Sec:Proposed}
Although the Nelder--Mead algorithm has been successfully demonstrated for system identification of stochastic oscillators in~\cite{BN:2017,LKGL:2021}, it suffers from some well-known drawbacks. For example, the convergence rate of the Nelder--Mead algorithm is known to be fairly slow; it was shown in~\cite{HN:2006} that the convergence rate deteriorates as the dimension of the unknown increases in general. Moreover, the algorithm may converge to a nonstationary point even if the cost function enjoys good properties such as strict convexity and differentiability~\cite{McKinnon:1998}.
%We note that the convergence of the Nelder--Mead algorithm to a minimizer has been proven for several restricted cases only; see, e.g.,~\cite{LRWW:1998,LPW:2012}.

In this section, we propose a new derivative-free algorithm for solving the minimization problem~\eqref{min} with better performance. To construct a tailored algorithm for~\eqref{min}, we first observe several particular properties of the minimization problem~\eqref{min} and its cost function~\eqref{cost}.

% Property: Nonlinear least squares
\begin{property}
\label{Prop:LS}
The minimization problem~\eqref{min} is a nonlinear least squares problem.
\end{property}

Clearly,~\eqref{min} can be regarded as a weighted nonlinear least squares problem that has three model parameters $\epsilon$, $\alpha$, and $d$, and $2N_{\tau} N_a$ observations $\hD_{\tau_j}^{(n)}(a_i)$, $1 \leq n \leq 2$, $1 \leq i \leq N_a$, $1 \leq j \leq N_{\tau}$. 
Hence, various numerical solvers such as Gauss--Newton-type methods, gradient methods, and direct search methods can be considered; see, e.g.,~\cite{Yuan:2011} for a survey on numerical solvers for nonlinear least squares problems.
To decide which one is most suitable for solving~\eqref{min}, we investigate further properties of the cost function~\eqref{cost}.

% Property: Energy evaluation
\begin{property}
\label{Prop:eval}
The computational cost for an evaluation of the cost function $E(\epsilon, \alpha, d)$ is high.
\end{property}

As shown in \cref{Alg:residual}, one has to solve $2N_{a} N_{\tau}$ initial-boundary value problems of the form~\eqref{AFP} numerically in a single evaluation of the residual $\brho (\epsilon, \alpha, d)$.
It means that the computational cost for an evaluation of the cost function~\eqref{cost} is considerably high.
Hence, for numerical efficiency, direct search methods such as the Nelder--Mead algorithm should be avoided since they require a huge number of evaluations of the cost function in general.
Instead, Gauss--Newton type methods that require the information of the first-order derivatives may be considered.

At each iteration of a Gauss--Newton type method for solving~\eqref{min}, one has to solve a linear system~(see~\eqref{direction_search}) regarding the Jacobian $\bJ (\epsilon, \alpha, d)$
% \begin{equation*}
% \bJ (\epsilon, \alpha, d) = \begin{bmatrix}
% \dfrac{\partial \brho}{\partial \epsilon} & \dfrac{\partial \brho}{\partial \alpha} & \dfrac{\partial \brho}{\partial d} 
% \end{bmatrix}
% \end{equation*}
of the residual $\brho (\epsilon, \alpha, d)$.
Thanks to the following property of the minimization problem~\eqref{min}, the computational cost for solving such a linear system becomes marginal.

% Property: Number of unknowns
\begin{property}
\label{Prop:number}
The number of unknown parameters in~\eqref{min} is much smaller than the number of observations for fitting.
\end{property}

\cref{Prop:number} is straightforward since $3$ is far less than $2 N_a N_{\tau}$.
The dimension of the Jacobian $\bJ = \bJ (\epsilon, \alpha, d)$ is $2N_a N_{\tau} \times 3$, so that the matrix $\bJ^T \bP \bJ$ arising in the direction search equation has the size $3 \times 3$.
Hence, the linear system can be solved in a fairly short time even if the number of observations is very large.
This observation suggests us to adopt a Gauss--Newton type method to solve~\eqref{min}.
In particular, the Levenberg--Marquardt algorithm~\cite{Marquardt:1963} is a good option since it is well-known to show the robust and fast convergence behavior compared to other Gauss--Newton type methods.
However, the following property of the residual $\brho (\epsilon, \alpha, d)$ makes the situation not so straightforward.

% Property: No Jacobian
\begin{property}
\label{Prop:Jacobian}
The Jacobian $\bJ (\epsilon, \alpha, d)$ of the residual $\brho (\epsilon, \alpha, d)$ has no closed-form formula.
\end{property}

Because the term $D_{\tau_j}^{(n)}(a_i; \epsilon, \alpha, d)$ in~\eqref{cost} relies on the solutions of initial-boundary problems of the form~\eqref{AFP}, neither a closed-form formula nor a computationally cheap algorithm is available for finding the Jacobian $\bJ$ accurately.
Hence, the vanilla Levenberg--Marquardt algorithm is not directly applicable to~\eqref{min}.

The proposed algorithm is a modification of the Levenberg--Marquardt algorithm for~\eqref{min} so that it does not require the first-order derivatives of the cost function $E(\epsilon, \alpha, d)$.
In the proposed algorithm, we consider replacing $\bJ$ in the Levenberg--Marquardt algorithm by a finite difference approximation $\tilde{\bJ} = \tilde{\bJ} (\epsilon, \alpha, d)$:
\begin{equation}
\label{Jacobian}
\tilde{\bJ}(\epsilon, \alpha, d) = \begin{bmatrix}
\dfrac{\brho(\epsilon + \depsilon, \alpha, d) - \brho(\epsilon, \alpha, d)}{\depsilon} &
\dfrac{\brho(\epsilon, \alpha + \dalpha, d) - \brho(\epsilon, \alpha, d)}{\dalpha} &
\dfrac{\brho(\epsilon, \alpha, d + \dd) - \brho(\epsilon, \alpha, d)}{\dd} 
\end{bmatrix},
\end{equation}
where $\depsilon$, $\dalpha$, and $\dd$ are nonzero real constants sufficiently near $0$; we set
\begin{equation*}
\depsilon = \max \left\{ \frac{| \epsilon |}{10}, 10^{-5} \right\}, \quad
\dalpha = \max \left\{ \frac{| \alpha |}{10}, 10^{-5} \right\}, \quad
\dd = \max \left\{ \frac{|d|}{10}, 10^{-5} \right\}
\end{equation*}
heuristically.
Note that a single evaluation of the approximate Jacobian $\tilde{\bJ} (\epsilon, \alpha, d)$ in~\eqref{Jacobian} requires four evaluations of the residual: $\brho (\epsilon, \alpha, d)$, $\brho (\epsilon + \depsilon, \alpha, d)$, $\brho (\epsilon, \alpha + \dalpha, d)$, and $\brho (\epsilon, \alpha, d + \dd)$.

Once the approximate Jacobian $\tilde{\bJ}$ has been obtained, the search direction $\dtheta$ is computed by solving the following linear system:
\begin{equation}
\label{direction_search}
\left( \tilde{\bJ}^T \bP \tilde{\bJ} + \lambda \diag \left( \tilde{\bJ}^T \bP \tilde{\bJ} \right) \right) \dtheta = \tilde{\bJ}^T \bP \brho (\theta),
\end{equation}
where $\theta = [\epsilon, \alpha, d]^T$ and $\lambda$ is a positive damping factor adjusted at each iteration by the Marquardt rule~\cite{Marquardt:1963}.
%and $\tilde{\bM}$ is the damping matrix given by
%\begin{equation}
%\label{M}
%\tilde{\bM} = \diag \left( \tilde{\bJ}^T \bP \tilde{\bJ} %\right).
%\end{equation}
An initial value $\lambda_0$ for the damping factor $\lambda$ does not affect to the convergence behavior, and we simply set $\lambda_0 = 1$.

% Algorithm: Proposed algorithm
\begin{algorithm}
\caption{Proposed algorithm for solving~\eqref{min}}
\begin{algorithmic}[]
\label{Alg:proposed}
\STATE \textsc{Inputs}: system state $x(t)$, set of amplitudes $\{ a_i \}_{i=1}^{N_a}$, set of time-shifts $\{ \tau_{j} \}_{j=1}^{N_{\tau}}$
\STATE \textsc{Outputs}: parameters $\epsilon$, $\alpha$, and $d$

\STATE $\bullet$ Compute $\hD_{\tau_j}^{(n)}(a_i)$ from $x(t)$ using~\eqref{D_tau_exact}. \quad ($1 \leq n \leq 2$, $1 \leq i \leq N_a$, $1 \leq j \leq N_{\tau}$)
\STATE $\bullet$ Set the initial guess $(\epsilon^{(0)}, \alpha^{(0)}, d^{(0)})$ by the output of \cref{Alg:extrap}.

\FOR{$k=0,1,2,\dots$}
\STATE $\bullet$ Compute the approximate Jacobian $\tilde{\bJ} = \tilde{\bJ}(\theta^{(k)})$ by invoking the formula~\eqref{Jacobian}.
%\STATE $\bullet$ Compute the damping matrix $\tilde{\bM} = \tilde{\bM}(\theta^{(k)})$ by invoking the formula~\eqref{M}.
\STATE $\bullet$ Solve the linear system~\eqref{direction_search} with $\lambda = \lambda_k$ to find a candidate $\dtheta_0$ for the search direction.
\STATE $\bullet$ Solve the linear system~\eqref{direction_search} with $\lambda = \lambda_k/2$ to find a candidate $\dtheta_{-1}$ for the search direction.

\IF{$E(\theta^{(k)} - \dtheta_0) > E(\theta^{(k)})$ and $E(\theta^{(k)} - \dtheta_{-1}) > E(\theta^{(k)})$}
\STATE $\bullet$ Find the smallest $m > 0$ such that $E(\theta^{(k)} - \dtheta_m) \leq E(\theta^{(k)})$, where $\dtheta_m$ is the solution of~\eqref{direction_search} with $\lambda = 2^m \lambda_k$.
\STATE $\bullet$ Set $\theta^{(k+1)} = \theta^{(k)} - \dtheta_m$ and $\lambda_{k+1} = 2^m \lambda_k$.
\ELSE
\IF{$E(\theta^{(k)} - \dtheta_0) \leq E(\theta^{(k)} - \dtheta_{-1})$}
\STATE $\bullet$ Set $\theta^{(k+1)} = \theta^{(k)} - \dtheta_0$ and $\lambda_{k+1} = \lambda_k$. 
\ELSE
\STATE $\bullet$ Set $\theta^{(k+1)} = \theta^{(k)} - \dtheta_{-1}$ and $\lambda_{k+1} = \lambda_k / 2$. 
\ENDIF
\ENDIF
\STATE $\bullet$ Check the stop criteria.
\ENDFOR
\end{algorithmic}
\end{algorithm}

Finally, we summarize the proposed algorithm in \cref{Alg:proposed}.
It is well-known that Gauss--Newton type methods require a good initial guess for the unknown $\theta$ to ensure good performance~\cite{AK:2018}.
In the proposed method, we obtain a reliable initial guess $\theta^{(0)}$ by the extrapolation method~(see \cref{Alg:extrap}).
Note that the main computational cost of \cref{Alg:extrap} comes from two linear least-square problems~\eqref{LS1} and~\eqref{LS2}, so it is minor compared to the computational cost of a single iteration of the proposed algorithm.
Efficient implementation of \cref{Alg:proposed} on a multiprocessing computer system is discussed in \ref{App:Implementation}.

% Section: Demonstration and validation
\section{Demonstration and validation of the system-identification framework}
\label{Sec:Demonstration}
% Section: Synthetic data
\subsection{Synthetic data}
% Figure: epsilon experiment
\begin{figure}
\centering
\includegraphics[width=0.95\textwidth]{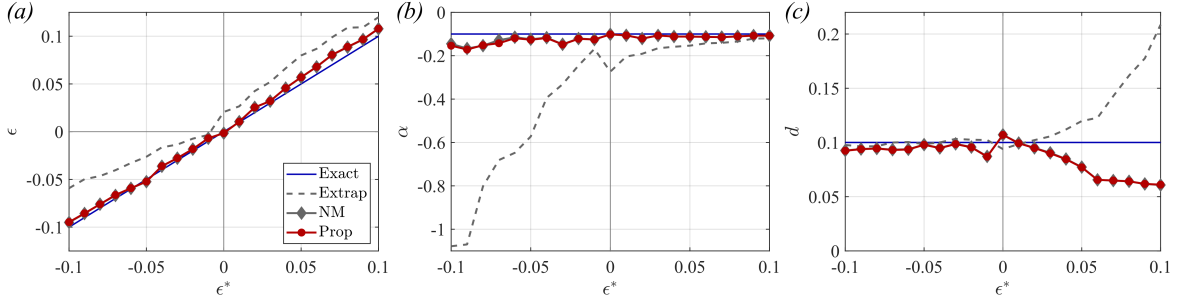}
\caption{Numerical results of system identification using three different algorithms: Extrap~(\cref{Alg:extrap}), NM~(Nelder--Mead algorithm), and Prop~(\cref{Alg:proposed}). Exact values for the parameters in the Van der Pol oscillator~\eqref{VdP} are chosen as $-0.1 \leq \epsilon^* \leq 0.1$, $\alpha^* = -0.1$, $d^* = 0.1$, and $\omega^* = 2\pi$. All values presented in this figure are averaged over 50 independent trials with respect to random noise $\eta$ in~\eqref{VdP}. Blue lines denote the true values of the input parameters $\epsilon$, $\alpha$, and $d$ in~\eqref{VdP}.}
\label{Fig:epsilon}
\end{figure}

% Figure: alpha experiment
\begin{figure}
\centering
\includegraphics[width=0.95\textwidth]{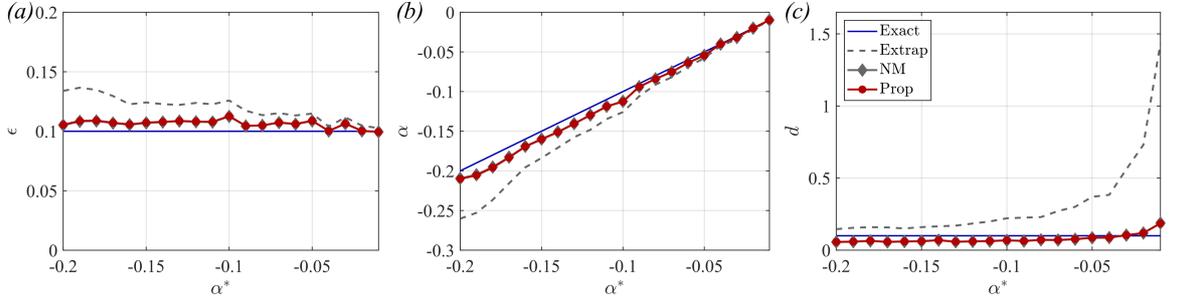}
\caption{Numerical results of system identification using three different algorithms: Extrap~(\cref{Alg:extrap}), NM~(Nelder--Mead algorithm), and Prop~(\cref{Alg:proposed}). Exact values for the parameters in the Van der Pol oscillator~\eqref{VdP} are chosen as $\epsilon^* = 0.1$, $-0.2 \leq \alpha^* \leq -0.01$, and $d^* = 0.1$. All values presented in this figure are averaged over 50 independent trials with respect to random noise $\eta$ in~\eqref{VdP}. Blue lines denote the true values of the input parameters $\epsilon$, $\alpha$, and $d$ in~\eqref{VdP}.}
\label{Fig:alpha}
\end{figure}

In order to demonstrate the performance of the proposed algorithm, we first conduct numerical simulations on synthetic data. A numerical Van der Pol oscillator perturbed by Gaussian white noise is chosen as the target system, and the system identification is repeatedly performed with varying linear~($-0.1 \leq \epsilon \leq +0.1$) and nonlinear~($-0.2 \leq \alpha \leq -0.01$) coefficients, with an interval of $0.01$ each. $d$, $\omega$, $N_a$ and $N_\tau$ are fixed at $0.1$, $2\pi$, $50$ and $100$, respectively. 
The set of amplitudes $\{ a_i \}_{i=1}^{N_a}$ is chosen as the equipartitioning points of the range of $a(t)$.
%, i.e.,
% \begin{equation*}
% a_i = a_{\min} + \frac{i-1}{N_a - 1}(a_{\max} - a_{\min}),
% \quad 1 \leq i \leq N_a.
% \end{equation*}
As for the time-delay input, we select $\tau_j$ so that the time-lagged signal has significant correlation with the original signal. That is, the set of time-delay values $\{ \tau_j \}_{j=1}^{N_{\tau}}$ is selected from the equipartitioned interval $[\tau_1, \tau_2]$, where the lower bound $\tau_1$ is the time lag that makes the autocorrelation of the signal drops below $0.97$ for $\epsilon>0$ and $0.6$ for $\epsilon \leq 0$. The upper bound $\tau_2$ is chosen to be 100 times the lower bound $\tau_1$.

In order to compensate the randomness of the experiments, each case is repeated 50 times with different random seeds while the results are averaged. We compare the results of extrapolation-based output-only system identification without optimization, existing output-only system identification combined with Nelder--Mead optimization \cite{BN:2017}, and the proposed system-identification framework introduced in \cref{Sec:Proposed}. Stop criteria for the optimization are described below and are applied for both the Nelder--Mead algorithm and the proposed algorithm:
\begin{equation*}
%\label{stop}
\frac{\left\| \theta^{(n+1)} - \theta^{(n)} \right\|_{\ell^2}}{1 + \left\| \theta^{(n)} \right\|_{\ell^2} } < 10^{-4}
\quad\textrm{and}\quad
\frac{\left| E(\theta^{(n+1)}) - E(\theta^{(n)}) \right|}{1 + \left| E(\theta^{(n)}) \right| } < 10^{-4}.
\end{equation*}
%which are the default stop criteria of the \texttt{fminsearch} program in MATLAB, a built-in implementation of the Nelder--Mead algorithm for unconstrained multivariable optimization.
MATLAB \texttt{pdepe} program is used as a numerical solver for the adjoint Fokker--Planck equation~\eqref{AFP}, in which the spatial and time discretizations are done by the method proposed in~\cite{SB:1990} and the \texttt{ode15s} program described in~\cite{SR:1997}, respectively. All codes are programmed using MATLAB R2021a and performed on a computer equipped with two Intel Xeon Gold 6240R CPUs~(2.4GHz, 24C), 192GB RAM, and the operating system CentOS~7.8 64-bit. 

\cref{Fig:epsilon,Fig:alpha} show results of system identification performed to the systems with varying $\epsilon$ and $\alpha$, respectively.
%(see \cref{Table:epsilon,Table:alpha} for tabulated results).
When the linear coefficient $\epsilon$ is varied, it can be found from \cref{Fig:epsilon}(a) that the extrapolation-based system identification gives a reasonable result, especially when $|\epsilon^*|$ is small. However, identified nonlinear coefficient $\alpha$ and the noise amplitude $d$ are found to be inaccurate when such a method is applied, specifically in the case where $\epsilon^*$ is either too small~(\cref{Fig:epsilon}(b)) or too large~(\cref{Fig:epsilon}(c)). One of the factors that cause this inaccuracy is the aforementioned finite-time effect arising from the small $\tau$ region. When this adverse effect is relieved with the optimization scheme, system-identification accuracy for $\epsilon$, $\alpha$, and $d$ dramatically increase. Interestingly, the accuracy of the existing method that uses the Nelder--Mead algorithm and the proposed algorithm has minimal differences. That is, we conclude that both the Nelder--Mead and proposed algorithms solve the minimization problem~\eqref{min} accurately and find the same local minimum. The advantage of the proposed algorithm, however, will soon become clear.

Similar trends are found when the nonlinear coefficient $\alpha$ is varied. In \cref{Fig:alpha}, extrapolation-based system identification yields an understandable results for $\epsilon$ and $\alpha$, but the error is greater for $d$ with an exponential growth of identified $d$ at large $\alpha$ region (see \cref{Fig:alpha}(c)). From either the Nelder--Mead optimization or the proposed algorithm, accurate values of $\epsilon$, $\alpha$, and $d$ are obtained.

% Figure: Energy decay
\begin{figure}
\centering
\includegraphics[width=0.65\textwidth]{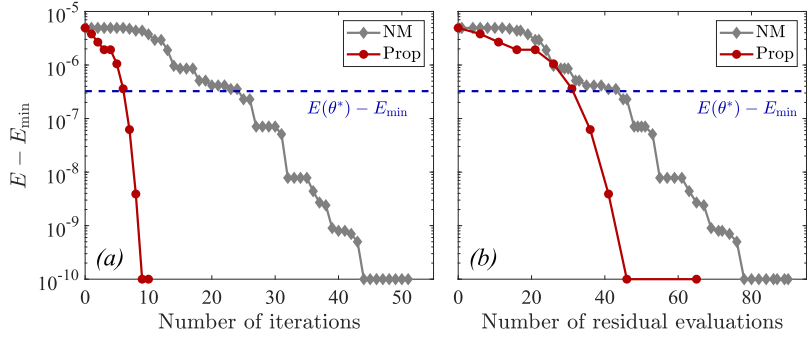}
\caption{Decay of the energy error $E(\theta^{(n)}) - E_{\min}$ for the Nelder--Mead algorithm~(NM) and \cref{Alg:proposed}~(Prop) with respect to \textbf{(a)}~the number of iterations and \textbf{(b)} the number of residual evaluations, where $E_{\min}$ is the energy value at the converged point~(both algorithms converge to the same point).
The blue dashed line indicates $E(\theta^*) - E_{\min}$, where $\theta^* = (\epsilon^*, \alpha^*, d^*) = (0.1, -0.1, 0.1)$ denotes the exact values for the parameters.}
\label{Fig:energy}
\end{figure}

% Figure: Convergence curves
\begin{figure}
\centering
\includegraphics[width=0.85\textwidth]{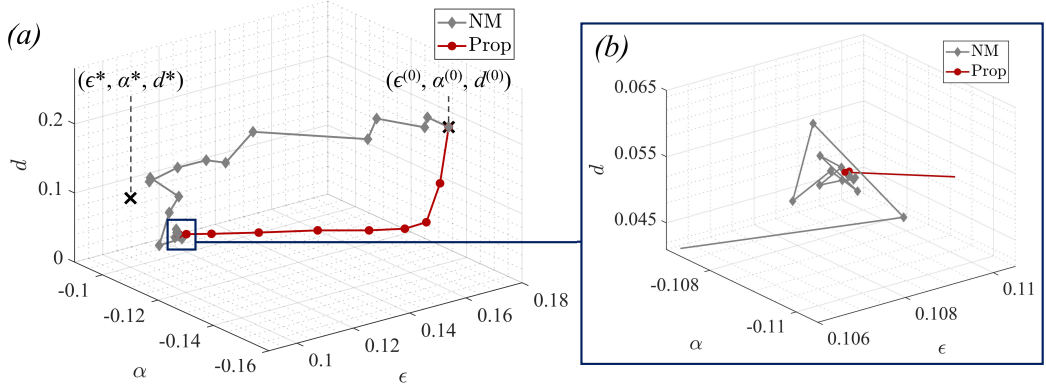}
\caption{Trajectories of the sequences $\{ (\epsilon^{(n)}, \alpha^{(n)}, d^{(n)}) \}$ generated by two algorithms NM~(Nelder--Mead algorithm) and Prop~(\cref{Alg:proposed}) plotted in the three-dimensional $(\epsilon, \alpha, d)$-space.
The red point $(\epsilon^*, \alpha^*, d^*) = (0.1, -0.1, 0.1)$ indicates the exact values for the parameters, and the blue point $(\epsilon^{(0)}, \alpha^{(0)}, d^{(0)})$ indicates the initial guess obtained by \cref{Alg:extrap}.}
\label{Fig:conv_curves}
\end{figure}

We now inspect the computational efficiency of the proposed system-identification method. \cref{Fig:energy} is the convergence plot showing the decay of the energy error $E(\theta^{(k)}) - E_{\min}$ with respect to the number of iterations $k$, where $E_{\min}$ is the cost value at the converged point. It can be found that, although the existing method (Nelder--Mead based optimization) and the proposed algorithm~(\cref{Alg:proposed}) reach a similar cost error, the latter requires a much lower number of iterations and residual evaluations. This means that the system identification can be performed much faster using the proposed algorithm, while its accuracy is maintained. 
The numerical efficiency of the proposed algorithm is further displayed in \cref{Fig:conv_curves}. The convergence trajectory shows that the coefficients are seamlessly converged to the target value, showing that our tailored optimization scheme works excellently for the system identification. %This is also evidenced by the minimal number of residual evaluations required for the proposed system-identification framework shown in \cref{Table:epsilon,Table:alpha}.

% Subsection: Gas-turbine combustor data
\subsection{Gas-turbine combustor data}
\begin{figure}
\centering
\includegraphics[trim={0.5cm 4cm 0.5cm 3.5cm}, clip, width=0.95\textwidth]{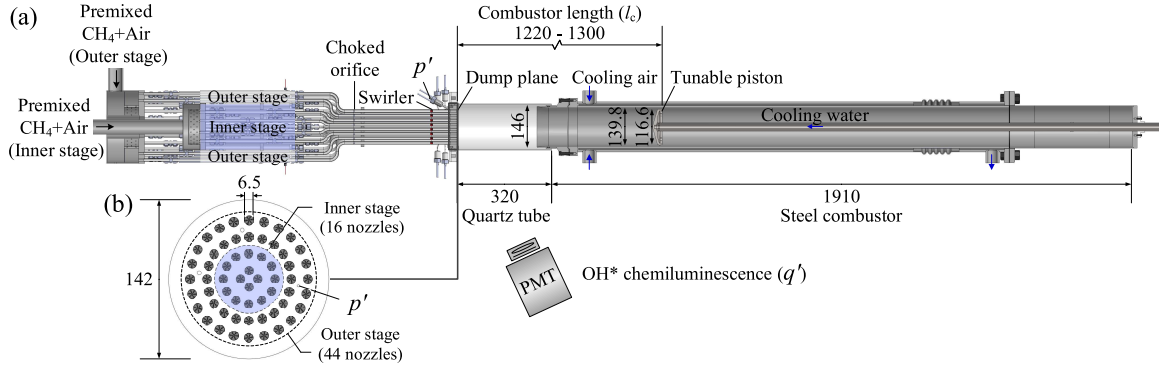}
\caption{Experimental setup of a mesoscale multi-nozzle model gas-turbine combustor with variable length for acquiring data used in this study, of which the cross-sectional view is shown in (a). This setup constitutes sixty swirl injectors, whose spatial arrangement at the combustor dump plane is shown in (b). At the combustor dump plane, acoustic pressure fluctuation ($p^{\prime}$) of the combustor is measured with a piezoelectric transducer. All dimensions are in millimeters. This figure is identical to \cite[Figure~1]{lee2022}.}
\label{Fig:setup}
\end{figure}

\begin{figure}
\centering
\includegraphics[width=0.9\textwidth]{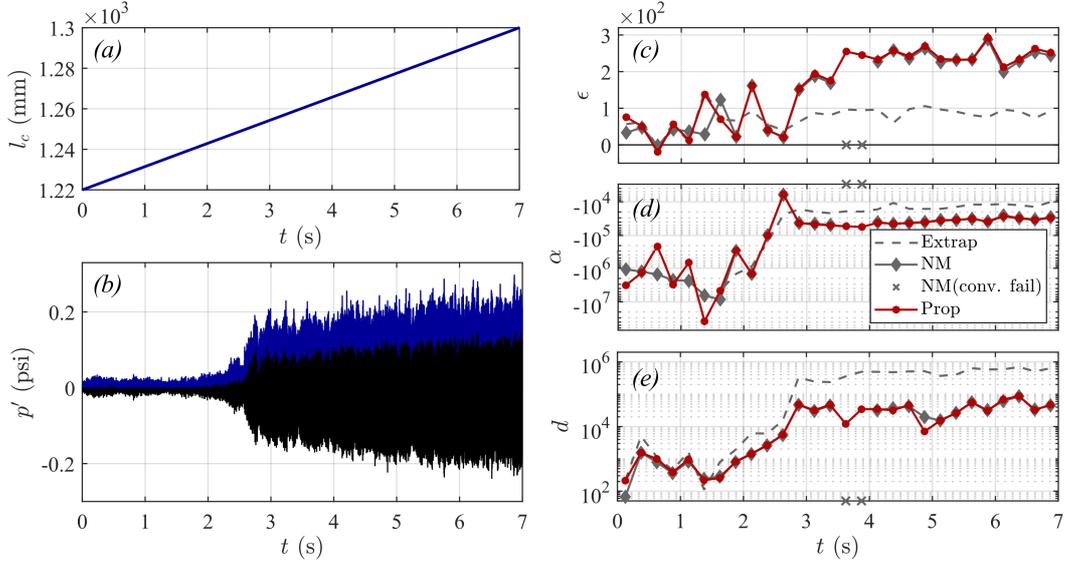}
\caption{(a) Change in combustor length and the corresponding (b) time-series signal of the acoustic pressure fluctuation (black line) and its amplitude obtained from the Hilbert transform (blue line). (c) Linear ($\epsilon$), (d) nonlinear ($\alpha$) coefficients, and (e) noise amplitude ($d$) obtained from the output-only system identification based on extrapolation (gray dashed lines), Nelder--Mead algorithm (gray markers), and the proposed method (red markers). Cross markers indicate the convergence failure for Nelder--Mead optimization.}
\label{Fig:expres}
\end{figure}

To further validate the developed framework on a realistic physical system, we use the data obtained from a model gas-turbine data identical to~\cite{lee2022}. This data features a transient development from the combustion noise to thermoacoustic instability of a combustor containing the methane-air premixed flame. A detailed description of the experimental setup used to acquire this data can be found in~\cite{lee2022,kang2021}. In brief, this setup shown in \cref{Fig:setup} enables the generation of thermoacoustic limit-cycle by adjusting the combustor length with the tunable piston. While varying the combustor length, pressure fluctuation $p^{\prime}$, which is treated as a system variable $x$ in equation~\eqref{VdP}, is measured with a piezoelectric transducer~(PCB 112A22, 14.5 mV/kPa).

By using a motorized traverse with a constant speed, the tunable piston is swept from 1220~mm to 1300~mm. Meanwhile, pressure fluctuation data is obtained with a sampling rate of 12~kHz for 7~seconds. The time evolution of $p^{\prime}$ and its amplitude during the sweep are shown in \cref{Fig:expres}(b). The time-series signal indicates that the system transitions from the low-amplitude oscillation to the high-amplitude limit-cycle oscillation at $t\approx2.5$s ($l_c\approx1250$). We slice this time-series data into 28 pieces~(0.25~seconds per segment) and conduct output-only system identification for each one-second segment. System identification results displayed in \cref{Fig:expres}(c, d) show the trends of increasing $\epsilon$, decreasing (and saturating) $\alpha$ upon the increase of combustor length $l_c$. We also find in \cref{Fig:expres}(e) that the amplitude of noise increases and saturates during the transition from stable fixed-point regime to the limit-cycle regime. These tendencies correctly capture the dynamics of the increasing fluctuation amplitude of the system variable $p^{\prime}$.

% Table: alpha experiment
\begin{table} \centering
\resizebox{0.7\textwidth}{!}{
\begin{tabular}{ccccccccccccccc}
\hline
\rowcolor{Gray}
Segment No. & 1 & 2 & 3 & 4 & 5 & 6 & 7 & 8 & 9 & 10 & 11 & 12 & 13 & 14 \\
\hline
NM & 332 & 279 & 300 & 258 & 310 & 295 & 319 & 281 & 291 & 220 & 231 & 275 & 268 & 253\\
Prop & 21 & 21 & 16 & 19 & 18 & 16 & 16 & 131 & 118 & 71 & 63 & 78 & 74 & 74\\
\hline \\[0.1cm] \hline

\hline
\rowcolor{Gray}
Segment No. & 15 & 16 & 17 & 18 & 19 & 20 & 21 & 22 & 23 & 24 & 25 & 26 & 27 & 28 \\
\hline
NM & N/A & N/A & 286 & 277 & 331 & 365 & 292 & 279 & 312 & 338 & 315 & 327 & 335 & 278\\
Prop & 92 & 130 & 82 & 88 & 94 & 101 & 136 & 91 & 88 & 93 & 89 & 106 & 90 & 91 \\
\hline
\end{tabular}
}
\caption{Number of residual evaluations in the Nelder--Mead algorithm~(NM) and the proposed method~(Prop) for system identification with gas-turbine combustor data.
N/A means that the method is not convergent.
Each segment corresponds to a time-series data of 0.25~seconds.}
\label{Table:residual}
\end{table}

To compare the computational efficiency of the Nelder--Mead and proposed algorithms, we present the number of residual evaluations in \cref{Table:residual}.
We observe that the numbers of the proposed method are consistently much lower than those of the Nelder--Mead algorithm.
These results verify that the proposed method outperforms the Nelder--Mead algorithm in terms of computational efficiency for system identification, even when dealing with realistic data.

Notably, we find that the proposed method can compute the system parameters at $t=$ 3.5--3.75~s, where the convergence of the Nelder--Mead algorithm has failed and the parameters could not be extracted. This is due to the poor optimization of the Nelder-Mead trajectory~(cf.~\cref{Fig:conv_curves}), which causes it to fall into a regime where the underlying adjoint Fokker-Planck equation~\eqref{AFP} becomes extremely stiff, leading to the failure of the numerical solver \texttt{pdepe}. In contrast, the proposed algorithm successfully avoids this regime and produces reliable results.

\begin{figure}
\centering
\includegraphics[width=0.9\textwidth]{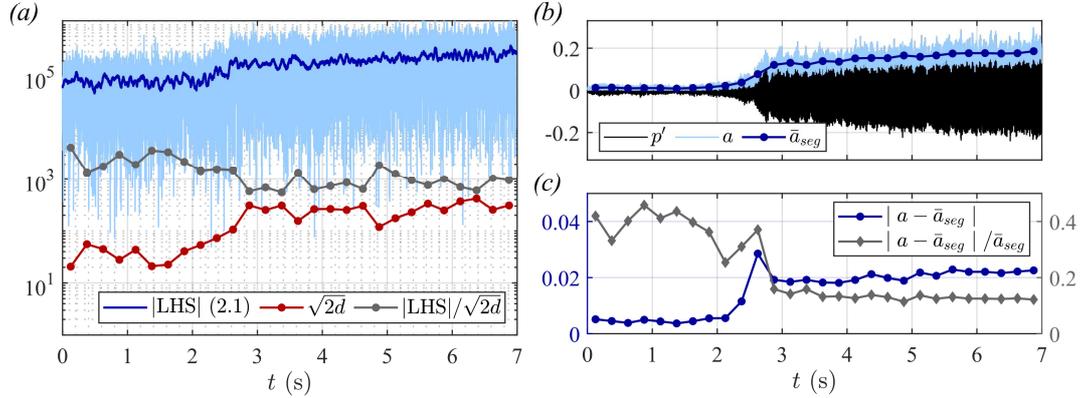}
\caption{(a) A comparison between scales of the left- and right-hand side of equation~\ref{VdP}. The light blue line indicates the absolute value of the left-hand side (LHS) of \eqref{VdP} computed with obtained coefficient, and the blue line is its average. The red line is the right-hand side of the same equation before the multiplication of Gaussian white noise. The grey line is the ratio between these two values. Y-axis is in the logarithmic scale. (b) The pressure fluctuation amplitude (light blue line) and its average in each data segment (blue points). (c) Deviation of pressure fluctuation amplitude due to the stochastic effects (blue points) and its ratio to the mean fluctuation amplitude in each segment (grey points).}
\label{Fig:noise}
\end{figure}

Finally, we compare the scales of deterministic and random parts of the noise-perturbed Van der Pol equation (see~\eqref{VdP}). We computed the absolute value of the deterministic part ($\mid$LHS$\mid$) and the scale of stochasticity ($\sqrt{2d}$) using the system parameters obtained from the proposed method. The results displayed in \cref{Fig:noise}(a) shows that, while both the deterministic and stochastic parts strengthen while the combustor length increases, the ratio between these two parts tends to decrease. This means that the system becomes more deterministic as the system approaches and enters the thermoacoustic limit cycle. During such a process, the ratio between $\mid$LHS$\mid$ and $\sqrt{2d}$ varies between 4200 and 550. This means that the noise acting on the system, regardless of its source, exerts 0.024--0.181\% of randomness to the deterministic second-derivative dynamics of the combustor. This amount of stochasticity is converted into 33--46\% deviation of the amplitude fluctuation in the stable fixed-point regime (0--1 s in the experiment; see \cref{Fig:noise}(b, c)), and 11--14\% deviation in the fully developed limit-cycle regime (5--7 s in the experiment). Thus, We can conclude that the proposed method can be successfully applied to the data affected by intense noise--including the inherent, external, and measurement noises--which perturbs the original deterministic signal for a maximum of 46\%.

% Section: Conclusion
\section{Conclusion}
\label{Sec:Conclusion}
In this paper, we proposed a numerically efficient system-identification framework for a stochastically driven self-sustained oscillator. In this framework, initial system parameters are extracted from the time-series data characterized by the Langevin equation and the corresponding Fokker--Planck equation. We then constructed an optimization model that extracts coefficients using the adjoint Fokker--Planck equation. We proposed a tailored algorithm for the optimization model, which is a derivative-free modification of the  Levenberg--Marquardt algorithm. Computational aspects of the proposed algorithm are considered to enhance the computation speed on a parallel computer. We demonstrated the proposed framework on both the numerical and experimental systems in the linearly stable state~($\epsilon<0$) and linearly unstable state~($\epsilon>0$) at self-sustained oscillation. As a result, we showed that the computational cost for identifying the system coefficients is greatly reduced without sacrificing the system-identification accuracy. From the analysis on stochasticity of the experimental data used in this study, we found that the proposed method can be applied to a highly noisy system.

A significant implication of this study is that the overall time required for performing the output-only system identification is substantially reduced. Considering that system identification is often required for the real-time monitoring and control of dynamical systems, the proposed framework opens up new possibilities for the instantaneous diagnosis and the feedback control of practical engineered systems.
%Although Gauss--Newton-type algorithms used in our study are known to be sensitive to the initial value, we found that the extrapolation of finite-time drift and diffusion terms can provide reliable initial values for the proposed algorithm. We can therefore conclude that the proposed framework can serve as a fast and accurate system-identification strategy. 

As for the limitation of this study, the proposed framework is only applicable to the system where the amplitude and phase of the oscillation change slowly. In other words, the absolute value of the growth rate of the system should be sufficiently low. A representative case where such a condition is met is found near a Hopf bifurcation. This implies that the proposed system-identification framework can nevertheless be used to diagnose or predict the instability features near a supercritical or subcritical Hopf bifurcation~\cite{lee_phd}.

% Authorship contribution statement
\section*{CRediT authorship contribution statement}
\textbf{Minwoo Lee}: Conceptualization, Validation, Investigation, Visualization, Writing.
\textbf{Kyu Tae Kim}: Validation, Investigation, Data curation.
\textbf{Jongho Park}: Methodology, Software, Investigation, Resources, Writing.

% Declaration of competing interest
\section*{Declaration of competing interest}
The authors declare that they have no known competing financial interests or personal relationships that could have appeared to influence the work reported in this paper.

% Acknowledgement
\section*{Acknowledgement}
Minwoo Lee was supported by the National Research Foundation of Korea~(NRF) grant funded by the Korean government~(MSIT)~(No.~2021R1G1A1091278). Kyu Tae Kim was supported by NRF grant funded by MSIT (No.~2022R1A2B5B01001554). Jongho Park's work was supported by NRF grant funded by MSIT~(No.~2021R1C1C2095193).

\appendix
% Appendix: VdP -> FP
\section{Derivation of the Fokker--Planck equation~\eqref{FP}}
\label{App:FP}
% Figure: Extraction of a(t) from x(t)
\begin{figure}
\centering
\includegraphics[width=0.75\hsize]{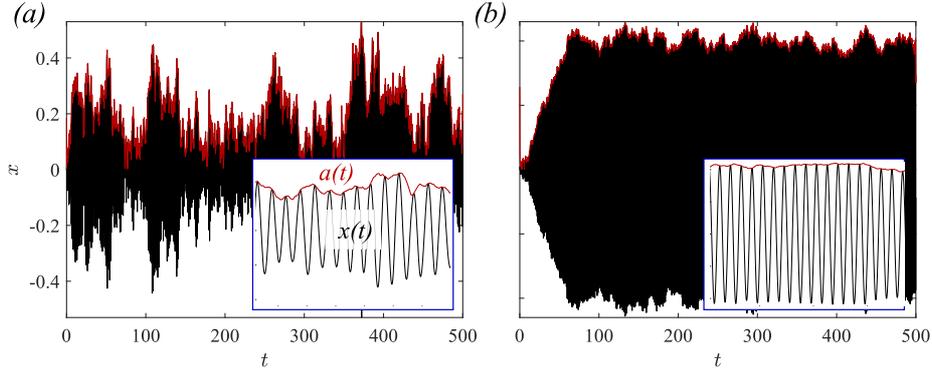}
\caption{The system state $x(t)$ and amplitude $a(t)$ of stochastically forced oscillators in \textbf{(a)} linearly stable state $(\epsilon, \alpha) = (-0.1, -0.1)$ and \textbf{(b)} linearly unstable state $(\epsilon, \alpha) = (0.1, -0.1)$. $\omega$ and $d$ are set to $2\pi$ and $0.1$, respectively.  The amplitude $a(t)$ is obtained from the Hilbert transform of $x(t)$. In both cases, $a(t)$ varies much slower than $x(t)$~(cf.~\cite[Figure~3]{lee_2019}).}
\label{Fig:extract_a}
\end{figure}

In this appendix, we provide a detailed derivation of the Fokker--Planck equation~\eqref{FP} from the Langevin equation~\eqref{VdP}. We first apply the method of variation of parameters~\cite{nayfeh1981} in~\eqref{VdP}. In other words, we transform the instantaneous state of the system $x(t)$ into its amplitude $a(t)$ and phase $\phi (t)$ as follows~\cite{roberts1986,ZHU1987421}:
\begin{align*}
x &= a(t) \cos \left( \omega t + \phi (t) \right), \\
\frac{dx}{dt} &= -a(t)\omega  \sin \left( \omega t + \phi (t) \right).
\end{align*}

An illustration of the signal $x(t)$ and the amplitude $a(t)$ is shown in \cref{Fig:extract_a}. From the substitution of variables and applying trigonometric identities, we obtain a system of ordinary differential equations with respect to $a (t)$ and $\phi (t)$ as follows~\cite{lee_2019}:
\begin{subequations}
\label{a_phi_system}
\begin{align}
\label{a_phi_system1}
\frac{da}{dt} &= \frac{\epsilon}{2}a + \frac{\alpha}{8}a^3 + Q_1 (\Phi) - \left( \frac{\sqrt{2d}}{\omega} \sin \Phi \right) \eta, \\
\label{a_phi_system2}
\frac{d\phi}{dt} &= Q_2 (\Phi) - \left( \frac{\sqrt{2d}}{\omega a} \cos \Phi \right) \eta,
\end{align}
\end{subequations}
where $\Phi (t) = \omega t + \phi (t)$, and $Q_1 (\Phi)$ and $Q_2 (\Phi)$ are the sum of all the terms with first-order cosine components appearing in~\eqref{a_phi_system1} and~\eqref{a_phi_system2}, respectively. By assuming that $a(t)$ and $\Phi (t)$ are slow variables (see \cref{Fig:extract_a}), $Q_1 (\Phi)$ and $Q_2 (\Phi)$ become zero upon time averaging. It is worth mentioning that, in the noise-free case~($d=0$),~\eqref{a_phi_system1} reduces to the normal-form equation of Hopf bifurcation:
\begin{equation*}
\frac{da}{dt} = \frac{\epsilon}{2}a + \frac{\alpha}{8}a^3.
\end{equation*}
For nonzero case, on the contrary, stochastic averaging~\cite{stratonovich1963,stratonovich1967} can be applied to~\eqref{a_phi_system}, yielding the following Fokker--Planck equation~\eqref{FP} whose unknown is the transitional probability density function $P(a,t)$ of $a$ at time $t$:
\begin{equation*} \begin{split}
&\frac{\partial}{\partial t} P (a,t) = - \frac{\partial}{\partial a} \left[ D^{(1)}(a) P(a,t) \right] + \frac{\partial^2}{\partial a^2} \left[ D^{(2)}(a) P(a,t) \right],
\quad a > 0, \gap t > 0, \\
&P(a,0) = P_0 (a), \quad a > 0, \\
&P(0, t) = 0, \quad t > 0.
\end{split} \end{equation*}
Note that the Kramer--Moyal coefficients $D^{(1)}(a)$ and $D^{(2)}(a)$ were given in~\eqref{KMc}.

% Appendix: Extrapolation method
\section{Extrapolation method for system identification}
\label{App:Extrapolation}
\begin{figure}
    \centering
    \includegraphics[width=0.85\hsize]{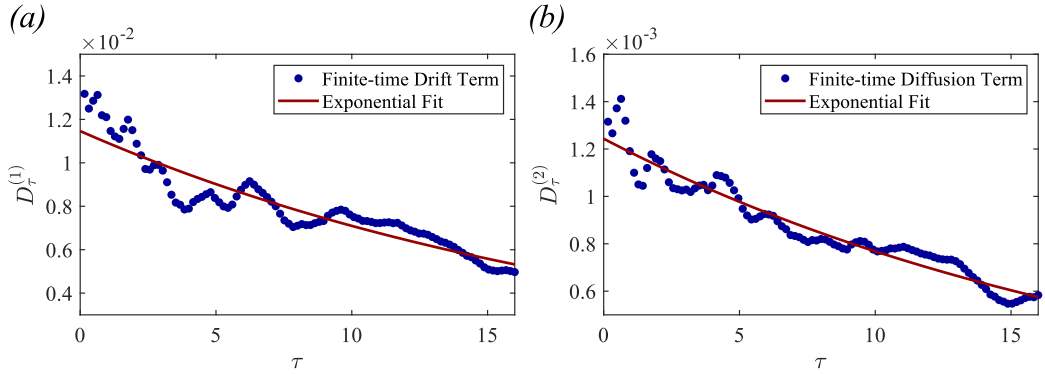}
    \caption{Experimental (a) Drift and (b) diffusion terms obtained from \eqref{D_tau_exact} and the corresponding extrapolation toward $\tau=0$ based on exponential fit \eqref{linear_regression}. The results are obtained at $\epsilon=0.1$, $\alpha=-0.1$, $d=0.1$, $\omega=2\pi$, and $a=1.86$.}
    \label{Fig:extrap}
\end{figure}

In this appendix, we present the extrapolation method for system identification of a self-sustained oscillator under stochastic forcing modeled by the Langevin equation~\eqref{VdP}.
Likewise the optimization method described in \cref{Sec:SI}, we estimate the Kramer--Moyal coefficients $D^{(1)} (a)$ and $D^{(2)} (a)$ in~\eqref{FP}.
We assume that the amplitude and time spaces are discretized as in \cref{Sec:SI}.
Namely, we have a set of amplitudes $\{ a_i \}_{i=1}^{N_a}$ and a set of time-shifts $\{ \tau_j \}_{j=1}^{N_{\tau}}$.
Recall that the estimate $\hD_{\tau_j}^{(n)} (a_i)$~($n = 1,2$) for $D_{\tau_j}^{(n)} (a_i)$ is computed by~\eqref{D_tau_exact}. In order to estimate $D^{(n)}(a)$ at zero time-delay limit~\eqref{D_tau_limit}, one may consider extrapolating finite-time drift and diffusion terms toward $\tau=0$, as depicted in \cref{Fig:extrap}. Specifically, we extrapolate $\hD_{\tau_j}^{(n)}(a_i)$ using an exponential fit~\cite{LKGL:2021}. For fixed $n$ and $a_i$, we assume that
\begin{equation*}
\hD_{\tau_j}^{(n)}(a_i) \approx \exp \left(c_1^{(n)} (a_i) \tau_j + c_0^{(n)} (a_i) \right)
\end{equation*}
for some constants $c_0^{(n)} (a_i)$ and $c_1^{(n)} (a_i)$. Equivalently, we have
\begin{equation}
\label{linear_regression}
\log \hD_{\tau_j}^{(n)}(a_i) \approx c_1^{(n)} (a_i) \tau + c_0^{(n)} (a_i).
\end{equation}
The least-squares solution of~\eqref{linear_regression} with respect to $\tau = \tau_1, \dots, \tau_{N_{\tau}}$ is given by
\begin{subequations}
\label{LS1}
\begin{equation}
\begin{bmatrix} c_0^{(n)} (a_i) \\ c_1^{(n)} (a_i) \end{bmatrix} = (\bA^T \bA)^{-1} \bA \bb,
\end{equation}
where
\begin{equation}
\bA = \begin{bmatrix} 1 & \tau_1 \\ \vdots & \vdots \\ 1 & \tau_{N_{\tau}} \end{bmatrix}, \quad
\bb = \begin{bmatrix} \log \hD_{\tau_1}^{(n)}(a_i) \\ \vdots \\ \log \hD_{\tau_{N_{\tau}}}^{(n)}(a_i) \end{bmatrix}.
\end{equation}
\end{subequations}
Note that $\bA^T \bA$ is invertible whenever all $\tau_j$'s are distinct.
The $\tau$-extrapolation of the diffusion term $\hD_{\tau}^{(2)} (a_i)$ toward $\tau = 0$ yields
\begin{equation*}
d = 2\omega^2 D^{(2)}(a_i) = 2 \omega^2 \lim_{\tau \rightarrow 0} D_{\tau}^{(2)}(a_i)
\approx 2 \omega^2 \exp \left( c_0^{(2)}(a_i) \right),
\end{equation*}
so that we may approximate $d$ by
\begin{equation}
\label{d_extrap}
d \approx \frac{2\omega^2}{N_a} \sum_{i=1}^{N_a} \exp \left( c_0^{(2)} (a_i) \right).
\end{equation}
Similarly, from the $\tau$-extrapolation of the drift term $\hD_{\tau}^{(1)} (a_i)$, we have
\begin{equation*}
\frac{\epsilon}{2}a_i + \frac{\alpha}{8}a_i^3 + \frac{d}{2\omega^2 a_i}
= D^{(1)}(a_i) = \lim_{\tau \rightarrow 0} D_{\tau}^{(1)}(a_i) \approx \exp \left( c_0^{(1)} (a_i) \right).
\end{equation*}
In order to identify $\epsilon$ and $\alpha$ from the above relation, we solve the following least-squares problem with respect to $a = a_1, \dots, a_{N_a}$:
\begin{subequations}
\label{LS2}
\begin{equation}
\begin{bmatrix} \epsilon \\ \alpha \end{bmatrix}
= (\widetilde{\bA}^T \widetilde{\bA})^{-1} \widetilde{\bA} \tilde{\bb},
\end{equation}
where
\begin{equation}
\widetilde{\bA} = \begin{bmatrix} \frac{a_1}{2} & \frac{a_1^3}{8} \\ \vdots & \vdots \\ \frac{a_{N_a}}{2} & \frac{a_{N_a}^3}{8} \end{bmatrix}, \quad
\tilde{\bb} = \begin{bmatrix} \exp \left(c_0^{(1)}(a_1) \right) - \frac{d}{2 \omega^2 a_1} \\ \vdots \\ \exp \left(c_0^{(1)}(a_{N_a}) \right) - \frac{d}{2 \omega^2 a_{N_a}} \end{bmatrix}.
\end{equation}
\end{subequations}
In~\eqref{LS2}, the matrix $\widetilde{\bA}^T \widetilde{\bA}$ is invertible if and only if all $a_i$'s are distinct.
We summarize the overall procedure to find the parameters $\epsilon$, $\alpha$, and $d$ by extrapolation in \cref{Alg:extrap}.

% Algorithm: Extrapolation
\begin{algorithm}
\caption{Extrapolation process for finding $D^{(n)}$}
\begin{algorithmic}[]
\label{Alg:extrap}
\STATE \textsc{Inputs}: system state $x(t)$, set of amplitudes $\{ a_i \}_{i=1}^{N_a}$, set of time-shifts $\{ \tau_{j} \}_{j=1}^{N_{\tau}}$
\STATE \textsc{Outputs}: parameters $\epsilon$, $\alpha$, and $d$

\STATE $\bullet$ Compute $\hD_{\tau_j}^{(n)}(a_i)$ from $x(t)$ by invoking the formula~\eqref{D_tau_exact}. \quad ($1 \leq n \leq 2$, $1 \leq i \leq N_a$, $1 \leq j \leq N_{\tau}$)
\STATE $\bullet$ Compute $c_0^{(n)}(a_i)$ and $c_1^{(n)}(a_i)$ by solving the least-squares problem~\eqref{LS1} with $a = a_i$. \quad ($1 \leq n \leq 2$, $1 \leq i \leq N_a$)
\STATE $\bullet$ Compute $d$ by invoking the formula~\eqref{d_extrap}.
\STATE $\bullet$ Compute $\epsilon$ and $\alpha$ by solving the least-squares problem~\eqref{LS2}.
\end{algorithmic}
\end{algorithm}

% Figure: Extrapolation process

\cref{Fig:extrap} shows numerically generated $D_{\tau}^{(n)}(a)$ at $a=1.86$ and their regression $\hD_{\tau_j}^{(n)}(a_i)$ based on \eqref{linear_regression}. Although the regression and the corresponding extrapolation reasonably represents the experimental $D_{\tau}^{(n)}(a)$, there exist some gap between $D_{\tau}^{(n)}(a)$ and $\hD_{\tau_j}^{(n)}(a_i)$, especially at small $\tau$. This is because the finite-time effect that occurs due to the non-Markovian properties of the noise acting on the system comes into play. The finite-time effect becomes highly significant in the coarsely sampled data or the bandpass-filtered signal \cite{BN:2017}. Because of this effect, the extrapolation-based drift and diffusion terms and the system parameters obtained from them contain an inherent inaccuracy. Nevertheless, it can be found from \cref{Fig:extrap} that the exponential regression can reasonably reproduce the behavior of $D_{\tau}^{(n)}(a)$. Therefore, the system parameters obtained here can serve as a reliable initial guess for the proposed optimization-based system identification process proposed in this paper.

% Appendix: Implementation issues of the proposed algorithm
\section{Implementation issues of the proposed algorithm}
\label{App:Implementation}
The appendix addresses computational efficiency and implementation considerations for the proposed algorithm.
The computational efficiency of the algorithm is notably influenced by the total count of residual evaluations, as outlined in \cref{Prop:eval}. We examine the number of residual evaluations needed during each iteration of \cref{Alg:proposed} and explore efficient strategies for implementing the algorithm on a multiprocessing computer system.

At the $k$th iteration of \cref{Alg:proposed}, we require at least six values of the residual: at the points $\theta^{(k)}$, $\theta^{(k)} + (\depsilon, 0, 0)$, $\theta^{(k)} + (0, \dalpha, 0)$, $\theta^{(k)} + (0, 0, \dd)$, $\theta^{(k)} - \dtheta_0$, and $\theta^{(k)} - \dtheta_{-1}$.
Among them, $\brho (\theta^{(k)})$ need not to be evaluated because it has been already evaluated at the previous iteration and we are able to reuse it.
Three values $\theta^{(k)} + (\depsilon, 0, 0)$, $\theta^{(k)} + (0, \dalpha, 0)$, and $\theta^{(k)} + (0, 0, \dd)$ can be computed independently, hence they can be processed in parallel.
Similarly, $\theta^{(k)} - \dtheta_0$ and $\theta^{(k)} - \dtheta_{-1}$ can be evaluated in parallel.
Therefore, it requires approximately twice the residual evaluation time to evaluate the above-mentioned six residual values if parallel processing is efficiently utilized.

When both $E(\theta^{(k)} - \dtheta_0)$ and $E(\theta^{(k)} - \dtheta_{-1})$ are greater than $E(\theta^{(k)})$, we require $m$ additional evaluations of the residual at the points $\theta^{(k)} - \dtheta_1$, \dots, $\theta^{(k)} - \dtheta_m$.
Fortunately, it can be observed numerically that such a case seldom occurs and does not severely deteriorate the performance of the algorithm.
Indeed, the Levenberg--Marquardt algorithm is known to act more like the Gauss--Newton method, i.e., the damping factor $\lambda$ decreases and the term $\tilde{\bJ}^T \bP \tilde{\bJ}$ becomes more effective when the estimate $\theta^{(k)}$ becomes close to a local minimum of the cost function~\cite{Gavin:2019}.

% Figure: Flow chart for parallel computation
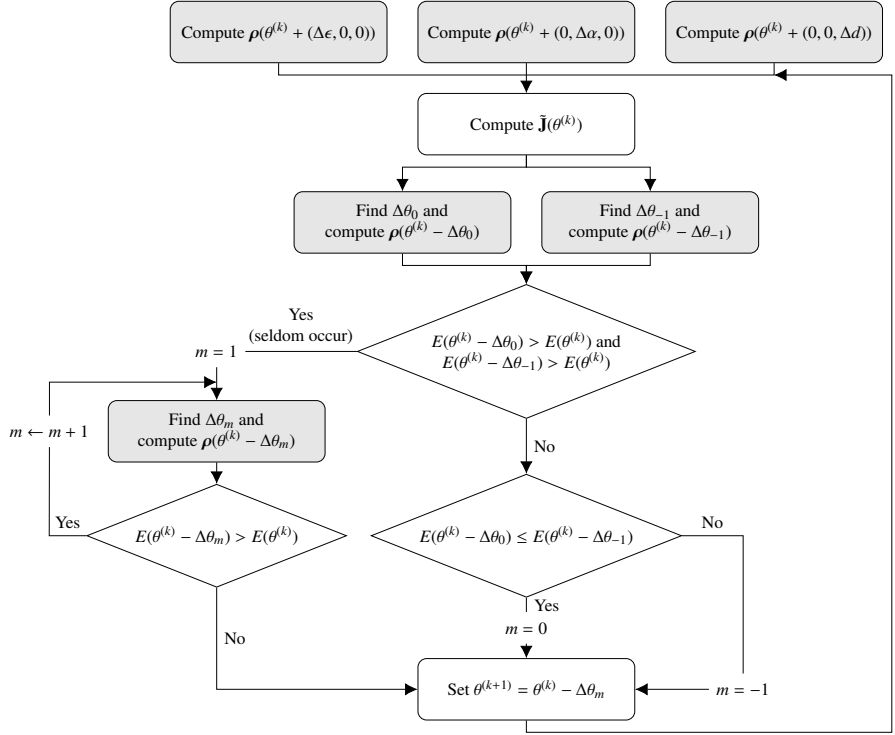
\begin{figure}
    \centering
    \resizebox{0.72\textwidth}{!}{
\begin{tikzpicture}[node distance=1.5cm,
    every node/.style={fill=white, font=\sffamily}, align=center]
  \tikzstyle{every node}=[font=\footnotesize]
 
  % Specification of nodes
  \node (FD1) [grayblock] {Compute $\brho (\theta^{(k)} + (\depsilon,0,0))$};
  \node (FD2) [grayblock, right of=FD1, xshift=2.5cm] {Compute $\brho (\theta^{(k)} + (0,\dalpha,0))$};
  \node (FD3) [grayblock, right of=FD2, xshift=2.5cm] {Compute $\brho (\theta^{(k)} + (0,0,\dd))$};
  \node (Jacobian) [block, below of=FD2] {Compute $\tilde{\bJ} (\theta^{(k)})$};
  \node (search0) [grayblock, below of=Jacobian, yshift=-0.1cm, xshift=-2cm] {Find $\dtheta_0$ and \\ compute $\brho(\theta^{(k)} - \dtheta_0)$};
  \node (search-1) [grayblock, below of=Jacobian, yshift=-0.1cm, xshift=2cm] {Find $\dtheta_{-1}$ and \\ compute $\brho(\theta^{(k)} - \dtheta_{-1})$};
  \node (comparison1) [ifblock, below of=Jacobian, yshift=-2.2cm] {$E(\theta^{(k)} - \dtheta_0) > E(\theta^{(k)})$ and \\ $E(\theta^{(k)} - \dtheta_{-1}) > E(\theta^{(k)})$};
  \node (m1) [nullblock, left of=comparison1, xshift=-3.5cm] {$m=1$};
  \node (searchm) [grayblock, below of=m1, yshift=0.2cm] {Find $\dtheta_{m}$ and \\ compute $\brho(\theta^{(k)} - \dtheta_{m})$};
  \node (comparison2) [ifblock, below of=searchm, yshift=-0.2cm] {$E(\theta^{(k)} - \dtheta_m) > E(\theta^{(k)})$};
  \node (m2) [nullblock, left of=searchm, xshift=-1.2cm] {$m \leftarrow m+1$};
  \node (comparison3) [ifblock, below of=comparison1, yshift=-1.5cm] {$E(\theta^{(k)} - \dtheta_0) \leq E(\theta^{(k)} - \dtheta_{-1})$};
  \node (m0) [nullblock, below of=comparison3] {$m=0$};
  \node (set) [block, below of=m0, yshift=0.5cm] {Set $\theta^{(k+1)} = \theta^{(k)} - \dtheta_m$};
  \node (m-1) [nullblock, right of=set, xshift=2cm] {$m=-1$};
  
  % Specification of lines between nodes
  \draw[-] (FD1) -- ++(0,-0.7) -- ++(4,0);
  \draw[->] (FD2) -- (Jacobian);
  \draw[-] (FD3) -- ++(0,-0.7) -- ++(-4,0);
  \draw[->] (Jacobian) -- ++(0,-0.7) -- ++(-2,0) -- (search0);
  \draw[->] (Jacobian) -- ++(0,-0.7) -- ++(2,0) -- (search-1);
  \draw[->] (search0) -- ++(0,-0.7) -- ++(2,0) -- (comparison1);
  \draw[->] (search-1) -- ++(0,-0.7) -- ++(-2,0) -- (comparison1);
  \draw[-] (comparison1) -- node [midway, above] {Yes \\ (seldom occur)} (m1);
  \draw[->] (m1) -- (searchm);
  \draw[->] (searchm) -- (comparison2);
  \draw[-] (comparison2.west) -- node [midway, above] {Yes} ++ (-0.6,0) -- (m2);
  \draw[->] (m2) -- ++(0,0.8) -- ++(2.7,0);
  \draw[->] (comparison1) -- node [midway, right] {No} (comparison3);
  \draw[-] (comparison3) -- node [midway, right] {Yes} (m0);
  \draw[->] (m0) -- (set);
  \draw[->] (comparison2) -- node [midway, right] {No} ++(0, -2.5) -- (set);
  \draw[-] (comparison3) -- node [midway, above] {No} ++(3.5,0) -- (m-1);
  \draw[->] (m-1) -- (set);
  \draw[->] (set) -- ++(0,-0.7) -- ++(5.9,0) -- ++(0,10.7) -- ++(-1.9,0);
\end{tikzpicture}
}
    \caption{Flow chart describing parallel implementation for the $k$th iteration of \cref{Alg:proposed}. Items positioned in the same row can be processed in parallel. Items corresponding to residual evaluations are depicted in gray.}   
\label{Fig:parallel}
\end{figure}

A flow chart for a single iteration of the proposed algorithm emphasizing the aspect of parallel computation is presented in \cref{Fig:parallel}.
In \cref{Fig:parallel}, items positioned in the same row are independent to each other, so that they can be processed simultaneously on a parallel computer.

% References
\bibliographystyle{elsarticle-num-names} 
\bibliography{refs_SI}

\end{document}